\documentclass[jkps,preprint,fleqn,showpacs,showkeys]{revtex4}
\usepackage{graphicx}
\usepackage{amssymb}
\usepackage{amsmath}
\usepackage{bm}
\begin{document}
\setcounter{page}{0}
\title[]{Dirac Phenomenological Analyses of Unpolarized Proton Inelastic Scattering from $^{22}$Ne}	
\author{ Moon-Won \surname{Kim} and Sugie \surname{Shim}}
\email{shim@kongju.ac.kr}
\thanks{Fax: +82-41-850-8489}
\affiliation{Department of Physics, Kongju National University, Gongju 314-701}

\date[]{Received 2014}

\begin{abstract}
Unpolarized 800 MeV proton inelastic scatterings from an s-d shell nucleus $^{22}$Ne are analyzed using phenomenological
optical potentials in the Dirac coupled channel formalism. The first-order rotational collective model is used to obtain the transition optical potentials for the low lying excited collective states that belong to the ground state rotational band of the nucleus. The optical potential parameters of Woods-Saxon shape and the deformation parameters of the excited states are varied phenomenologically using the sequential iteration method to reproduce the experimental differential cross section data. The effective central and spin-orbit optical potentials are obtained by reducing the Dirac equations to the Schr\"{o}dinger-like second-order differential equations and the surface-peaked phenomena are observed at the real effective central potentials when the scattering from $^{22}$Ne is considered. The obtained deformation parameters of the excited states are compared with those of the nonrelativistic calculations and another s-d shell nucleus $^{20}$Ne. The deformation parameters for the $2^+$ and the $4^+$ states of the ground state rotational band at the nucleus $^{22}$Ne are found to be smaller than those of $^{20}$Ne, indicating that the couplings of those states to the ground state are weaker at the nucleus $^{22}$Ne compared to those at the nucleus $^{20}$Ne. The multistep channel coupling effect is confirmed to be important for the $4^+$ state excitation of the ground state rotational band at the proton inelastic scattering from the s-d shell nucleus $^{22}$Ne.
\end{abstract}



\maketitle

\section{INTRODUCTION}

Relativistic analyses based on the Dirac equation have shown that they can achieve better agreement with experimental intermediate energy proton scattering data than the nonrelativistic analyses based on the Schr\"{o}dinger equation\cite{1,2,3}.
Because the Dirac analyses have proven to be very successful for the intermediate energy proton elastic scatterings from the spherically symmetric nuclei and a few deformed nuclei\cite{3,4,5,6}, the relativistic approaches have been expanded to the inelastic scatterings and have shown considerable improvements compared to the conventional nonrelativistic analyses\cite{7,8,9,10}. One of the merits of the Dirac approach instead of using the nonrelativistic approach is that the spin-orbit potential appears naturally in the Dirac approach when the Dirac equation is reduced to a Schr\"{o}dinger-like second-order differential equation, while the spin-orbit potential should be inserted by hand in the nonrelativistic Schr\"{o}dinger approach.

In this work we performed a relativistic Dirac analysis for the inelastic proton scatterings from an s-d shell nucleus $^{22}$Ne by using the optical potential model\cite{1} and the first-order collective model. Dirac phenomenological optical potentials in Woods-Saxon(W-S) shape are used, employing the scalar-vector (S-V) model where only Lorentz-covariant scalar and time-like vector optical potentials are included in the calculation.
The first-order collective rotational model is employed in order to describe the
collective motion of the excited states of the ground state rotational band(GSRB) in the nucleus.
A computer program called ECIS is used to solve the complicated Dirac coupled channel equations, where the Dirac optical potential and deformation parameters are determined phenomenologically using a sequential iteration method\cite{11}.  The Dirac equations are reduced to the Schr\"{o}dinger-like second-order differential equations to obtain the effective central and spin-orbit optical potentials and the results are analyzed and compared with those of another s-d shell nucleus $^{20}$Ne. The calculated results for the deformation parameters for the excited states of the ground state rotational band in the nucleus are analyzed and compared with those of $^{20}$Ne and the nonrelativistic approaches.

\section{Theory and Results}

Dirac Analyses are performed phenomenologically for the 800 MeV unpolarized proton inelastic scatterings from $^{22}$Ne by using optical potential model and the collective model.
Because $^{22}$Ne is one of the spin-0 nuclei,  only scalar,  time-like  vector  and  tensor
optical potentials  survive\cite{4, 12,13}, as  in spherically  symmetric nuclei\cite{14}; hence, the relevant Dirac equation for the elastic scattering from the nucleus is given as
\begin{equation}
[\alpha \cdot p + \beta ( m + U_S ) - ( E - U_V^0- V_c )
+ i \alpha \cdot  \hat{r} \beta U_T ] \Psi(r) = 0 .
\label{e1}
\end{equation}
Here, $U_S$ is a scalar potential, $U_V^0$ is a time-like vector potential, $U_T$ is a tensor potential, and $V_c $ is the Coulomb potential.
The scalar and time-like vector potentials are used as direct potentials in the
calculation, neglecting the tensor potentials since they have been found to  be always very
small compared  to scalar or vector potentials\cite{8, 10, 15} even though they are always present due to the
interaction  of the  anomalous magnetic  moment  of the  projectile with the
charge distribution of the  target. The scalar and vector optical potentials are complex and given as
\begin{equation}
U_S = V_S f_s (r) + i W_S g_s (r )
\label{e2}
\end{equation}
\begin{equation}
U_V^0 = V_V^0 f_v ( r) + i W_V^0 g_v ( r),
\label{e3}
\end{equation}
where $V_S$ and $W_S$ are the strengths of the real and the imaginary scalar potentials, $V_V^0$ and $W_V^0$ are the strengths of real and the imaginary time-like vector potentials, respectively. We assume that  these potentials have Fermi distribution as they are assumed to follow the distribution of nuclear density.
Fermi model form factors of Woods-Saxon shape for the Dirac optical potentials are given as
\begin{equation}
f_i (r ), g_i ( r) = {1 \over 1+ \rm{exp} \it{[{(r-R_{0,i}) \over Z_i } ]}},
\label{e4}
\end{equation}
where $R_{0,i}$ and $Z_i $ are potential radius and diffusiveness, respectively and the subscript $i$ stands for the real and imaginary scalar, and the real
and imaginary vector potentials.
In the first-order rotational model of ECIS, the deformation of the radius of the optical potential is given using the Legendre polynomial expansion method;
\begin{equation}
R(\theta ) = R_0 ( 1+ \beta_2 Y _{20 }+ \beta_4  Y _{40 }+ \cdots ),
\label{e5}
\end{equation}
with $R_0$ the radius at equilibrium, $\beta_\lambda$ is a deformation parameter and $\lambda$ is the multipolarity.
We assume that the shape of the deformed potentials follows the shape of the deformed nuclear densities and that the transition potentials can be obtained by assuming that they are proportional to the first-order derivatives of the diagonal potentials. However, depending on the  model assumed, pseudo-scalar and
axial-vector potentials may also be present in the equation when we consider inelastic
scattering. In the collective model approach used in this work, we assume that we can
obtain appropriate transition potentials by deforming the  direct potentials that describe the elastic
channel reasonably well\cite{14}.
In order to compare the calculated results with those of the previous nonrelativistic calculations, we reduce the Dirac equation to a Schr\"{o}dinger-like second-order differential equation by considering the upper component of the Dirac wave function to obtain the effective central and spin-orbit optical potentials\cite{3}.
The experimental data for the differential cross sections are obtained from Ref. 16 for the 800 MeV unpolarized proton inelastic scatterings from $^{22}$Ne.

As a first step, the 12 parameters of the diagonal scalar and vector potentials in Woods-Saxon shapes are determined by fitting the experimental elastic scattering data.
 The calculated results are shown as dash-dot lines in Fig. 1, and it is found that the observable elastic experimental differential cross section data are reproduced quite well.
The Dirac equations are phenomenologically solved to obtain the best fitting parameters to the experimental data by using the minimum chi-sq($\chi^2$) method.

\begin{figure}
\includegraphics[width=10.0cm]{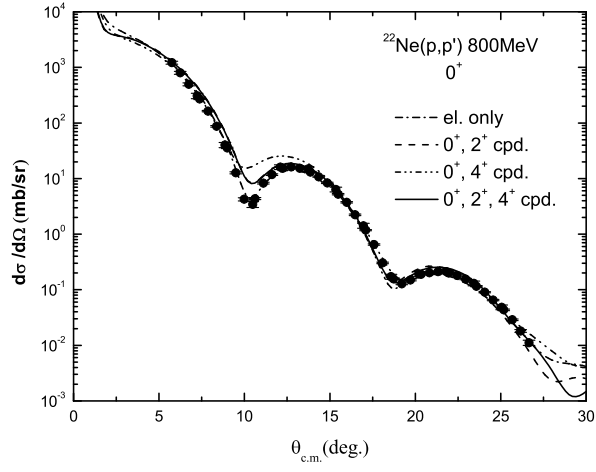}
\caption[0]{Differential cross section of the ground state for 800 MeV p +  $^{22}$Ne scattering. The dash-dot, dashed, dash-dot-dot and solid lines represent the results of Dirac phenomenological calculation where elastic scattering is considered, where the ground and the $2^+$ states are coupled, where the ground and the $4^+$ states are coupled, and where the ground, the $2^+$  and the $4^+$ states are coupled, respectively.}
\label{fig1}
\end{figure}

\begin{figure}
\includegraphics[width=10.0cm]{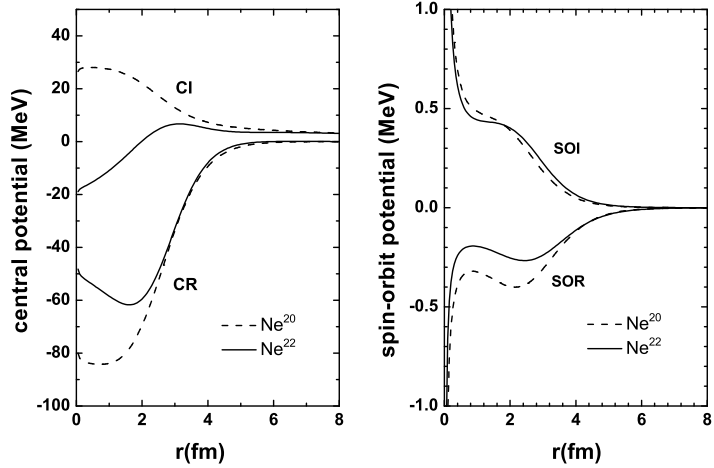}
\caption[0]{Comparison of the effective central and spin-orbit potentials of $^{22}$Ne  and $^{20}$Ne. CR and CI represent central real and imaginary potentials, and SOR and SOI represent spin-orbit real and imaginary optical
potentials, respectively.}
\label{fig2}

\end{figure}

\begin{table}
\caption{Calculated phenomenological optical potential parameters of a Woods-Saxon shape for 800 MeV proton elastic scatterings from $^{22}$Ne.}
\begin{ruledtabular}
\begin{tabular}{cccccccc}

   Potential      &   Strength (MeV)   & Radius (fm)  & Diffusiveness (fm)  ~ \\
   \hline
 Scalar  & -177.6     & 2.793 & 0.8065        ~ \\
 real    &      &     &            ~ \\ \hline
 Scalar  & 160.0     & 2.213 & 0.7792        ~ \\
 imaginary    &      &     &            ~ \\ \hline
 Vector  & 75.95    & 3.098 & 0.7176         ~ \\
 real    &      &     &           ~ \\ \hline
 Vector  & -101.6     & 2.649 & 0.6332       ~ \\
 imaginary  &      &     &             ~ \\
\end{tabular}
\end{ruledtabular}
\label{table1}
\end{table}

The calculated optical potential parameters of the Woods-Saxon shape for the 800 MeV proton elastic scatterings from $^{22}$Ne are shown in Table I. $\chi^2/N$, where $N$ is the number of experimental data, was about 4.3.
We observe that the real parts of the scalar potentials and the imaginary parts of the vector potentials turn out to be large and negative, and that the imaginary parts of the scalar potentials and the real parts of the vector potentials turn out to be large and positive, showing the same pattern as in the spherically symmetric nuclei\cite{3}.
In Fig. 2 we compared the effective central and spin-orbit potentials of $^{22}$Ne with those of $^{20}$Ne. It should be noted that one of the merits of the relativistic approach based on the Dirac equation instead of using the nonrelativistic approach based on the Schr\"{o}dinger equation is that the spin-orbit potential appears naturally in the Dirac approach when the Dirac equation is reduced to a Schr\"{o}dinger-like second-order differential equation, while the spin-orbit potential should be inserted by hand using Woods-Saxon shape in the nonrelativistic Schr\"{o}dinger approach.
Surface-peaked phenomena are clearly observed for the real parts of the effective central potentials (CR) at $^{22}$Ne. $^{20}$Ne and other s-d shell nuclei such as $^{24}$Mg\cite{8, 15} also showed the same surface-peaked phenomena, even though they are less clearly shown. The potential strength for the real central potential was about -50MeV at the center of the nucleus, showing large value compared to that of nonrelativistic calculations that was about -3.9 MeV\cite{16}. Somehow, the potential strength for the imaginary central potential turned out to be negative, about -20MeV, at the center of the nucleus $^{22}$Ne, while those of $^{20}$Ne and other s-d shell nuclei have positive values\cite{8, 15} and that of nonrelativistic calculations was about 49.1 MeV. However, the imaginary central potential strength at the surface area turned out to be positive as shown in the figure. The surface-peaked phenomena are clearly shown at the effective spin-orbit potentials, and the effective spin-orbit potential strengths of $^{22}$Ne turned out to be about the same order with those of nonrelativistic calculations\cite{16}. We should note that the surface-peaked phenomena never appear at the the nonrelativistic approaches since they use the Woods-Saxon shapes for both the central and spin-orbit potentials.

 \begin{figure}
\includegraphics[width=10.0cm]{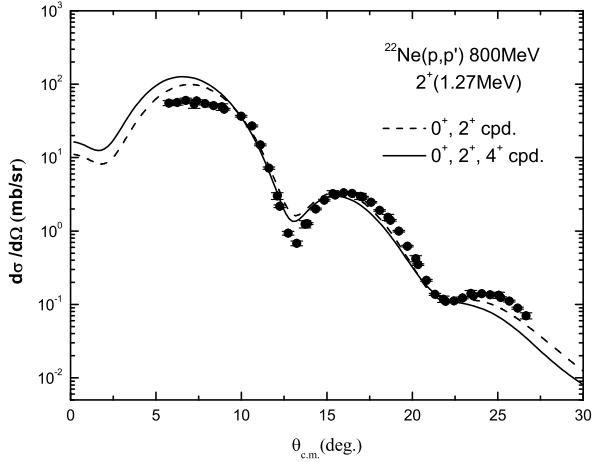}
\caption[0]{Differential  cross section of the $2^+ $ state for 800 MeV p +  $^{22}$Ne inelastic scattering. Dashed and solid lines represent the results of Dirac coupled channel calculation where the ground and the $2^+$ states are coupled and where the ground, the $2^+$ and the $4^+$ states are coupled, respectively.}
\label{fig3}
\end{figure}

\begin{figure}
\includegraphics[width=10.0cm]{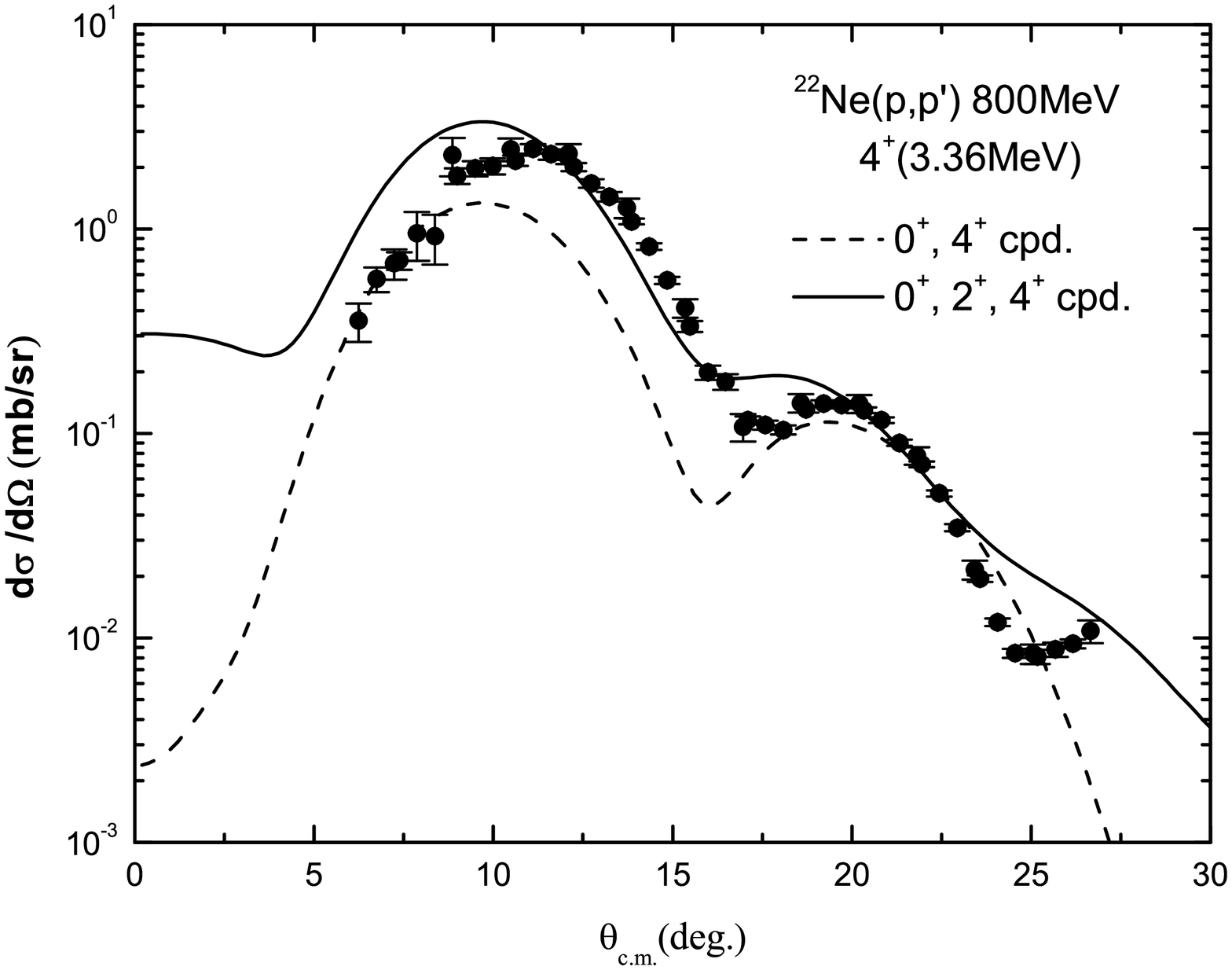}
\caption[0]{Differential  cross section of the $4^+ $ state for 800 MeV p +  $^{22}$Ne inelastic scattering. Dashed and solid lines represent the results of Dirac coupled channel calculation where the ground and the $4^+$ states are coupled and where the ground, the $2^+$ and the $4^+$ states are coupled, respectively.}
\label{fig4}
\end{figure}

Next, a six-parameter search is performed including one excited state, the $2^+$ or the $4^+$ state, in addition to the ground state, starting from the 12 parameters obtained for the direct optical potentials in the elastic scattering calculation. Here, the six parameters are the two deformation parameters, $\beta_S$ and $\beta_V$, of the excited state and the four potential strengths; the scalar real and imaginary potential strengths and the vector real and imaginary potential strengths, keeping the potential geometries unchanged. The optical potential strengths obtained by fitting the elastic scattering data in the elastic scattering calculation are varied because the channel coupling of the excited states to the ground state should be included in the inelastic scattering calculation.
Finally, an eight-parameter search is performed by considering all three states, the ground, the $2^+$ and the $4^+$ states, together in the calculation in order to investigate the effect of the channel coupling between the excited states and the results are compared with those of the calculation where only the ground and one excited states are coupled.
Figure 1 shows the results of the coupled channel calculations for the ground state and it is seen that the coupling effects with the excited states appears at the large angles, making the lines to go down\cite{3, 10}. In the figures, `cpd' means `coupled'. In Fig. 3 and 4, the calculated results for the the $2^+$ and the $4^+$ states are shown. For the $2^+$ state, the agreement with the experimental data didn't change much by including the coupling with $4^+$ state. $\chi^2/N$ for the two cases turned out to be about the same. However, the agreement with the experimental data for the $4^+$ state improved drastically by including the coupling with $2^+$ state, indicating two-step excitation via $2^+$ state is essential for the $4^+$ state excitation at the ground state rotational band. It means that it is essential to include the
multistep transition process because the low lying excited states
of the GSRB are strongly coupled each other, as shown in the inelastic scatterings from other axially-symmetric
deformed nuclei\cite{8, 9, 10, 15}.
This was not the case for the spherically
symmetric nuclei where the excited states
 could be well described by considering the coupling via
single-step  transitions\cite{3}. The potential strengths are changed to -311.8, 321.6, 125.0, and -155.0 MeV for scalar real and imaginary and vector real and imaginary potentials, respectively, in the $2^+$ state coupled case, -154.3, 302.3, 64.88, and -151.8 MeV in the $4^+$ state coupled case, and -293.1, 305.5, 118.4, and -150.3 MeV in the $2^+$ and $4^+$ states coupled case. These results confirm that the changes in the potential strengths depend on the coupling strength to the ground state; that is, the smallest change is seen for the $4^+$ state coupled case, and about the same change is seen for the $2^+$ state coupled case and  all three states coupled case.
The results of our relativistic coupled channel calculation showed slightly better agreement with experimental data than those of nonrelativistic calculations\cite{16}.

\begin{table}
\caption{Comparison of the deformation parameters for the $2^+ $ and the $4^+$ states for 800 MeV proton inelastic scatterings from $^{22}$Ne with those of $^{20}$Ne and nonrelativistic calculations.}
\begin{ruledtabular}
\begin{tabular}{c|ccccccc}
   &Target       &  Energy   &   &   &   ~ \\
   & nuclei      &  (MeV)  & $\beta_S $  & $\beta_V $  & $\beta_{NR} $  ~ \\
   \hline \hline
 $2^+ $ state  & $ ^{20}Ne $ & 1.63  &   .600   &  .587   &  $.47^{17} $       ~ \\ \cline{2-6}
 & $^{22}Ne $  & 1.27    & .202 & .313    &  $.46^{16}, .47^{19} $    ~ \\ \hline \hline
 $4^+ $ state  & $ ^{20}Ne $ & 4.25  &   .237   &  .245   &  $.25^{17} $       ~ \\ \cline{2-6}
  & $^{22}Ne $  & 3.36    & .019 & .048    &  $.10^{16}, .05^{19} $    ~ \\
\end{tabular}
\end{ruledtabular}
\label{table2}
\end{table}

In Table II, we show the deformation parameters for the $2^+$ and the $4^+$ states of $^{22}$Ne and $^{20}$Ne. It is shown that the deformation parameters for the $2^+$ and the $4^+$ states of $^{22}$Ne are smaller than those of $^{20}$Ne, even though the excitation energies for the states are smaller at the $^{22}$Ne. We can say that the couplings of the $2^+$ and the $4^+$ states to the ground state are weaker at the nucleus $^{22}$Ne compared to those at the nucleus $^{20}$Ne. The values obtained using the Dirac coupled channel calculations are also compared with those obtained by using the nonrelativistic coupled channel calculations\cite{16, 17, 18}. The obtained deformation parameters from Dirac phenomenological calculation for the $2^+$ and $4^+$ state excitations of $^{22}$Ne are found to have slightly smaller values compared to those of the nonrelativistic calculations.

\section{CONCLUSIONS}

A relativistic Dirac coupled channel calculation using an optical potential model could describe the low-lying excited states of the ground state rotational band for 800 MeV unpolarized proton inelastic scatterings from an s-d shell nucleus $^{22}$Ne reasonably well.
 The Dirac equations are reduced to second-order differential equations to obtain Schr\"{o}dinger-equivalent central and spin-orbit potentials, and surface-peaked phenomena were observed at the real effective central potentials for the scattering from $^{22}$Ne, as shown in the cases of $^{20}$Ne and $^{24}$Mg. The first-order rotational collective models are used to describe the low-lying excited states of the ground state rotational band in the nucleus, and the obtained deformation parameters are compared with those of  $^{20}$Ne. It is observed that the deformation parameters for the $2^+$ and the $4^+$ states of $^{22}$Ne are smaller than those of $^{20}$Ne, even though the excitation energies for the states are smaller at the nucleus $^{22}$Ne. We can say that the couplings of the $2^+$ and the $4^+$ states to the ground state are weaker at the nucleus $^{22}$Ne compared to those at the nucleus $^{20}$Ne.
 The deformation parameters for the  $2^+$ and the $4^+$ excited states are also compared with those of nonrelativistic calculations, and the obtained deformation parameters from the Dirac phenomenological calculation for the $2^+$ and $4^+$ state of $^{22}$Ne are found to have slightly smaller values compared to those of the nonrelativistic calculations.
 The multistep coupling effect is confirmed to be important for the $4^+$ state excitation of the ground state rotational band at the inelastic scattering from an s-d shell deformed nucleus $^{22}$Ne.

\begin{acknowledgments}
This work was supported by a research grant from Kongju National University in 2014.
\end{acknowledgments}


\begin{references}
\bibitem{1} L. G. Arnold, B. C. Clark, R. L. Mercer, and  P. Swandt, Phys. Rev. C {\bf 23}, 1949 (1981).
\bibitem{2} J. A. McNeil, J.  Shepard, and  S. J.  Wallace, Phys.  Rev. Lett  {\bf 50}, 1439 (1983); {\bf 50}, 1443 (1983).
\bibitem{3} S. Shim, Ph.D. dissertation, The Ohio State University 1989; L. Kurth, B. C. Clark, E. D. Cooper, S. Hama, S. Shim, R. L. Mercer, L. Ray, and G. W. Hoffmann, Phys.  Rev. C {\bf 49}, 2086 (1994).
\bibitem{4} S. Shim, B. C. Clark, E. D. Cooper, S. Hama, R. L. Mercer, L. Ray, J. Raynal, and H. S. Sherif, Phys. Rev. C {\bf 42}, 1592 (1990).
\bibitem{5} R. de Swiniarski, D. L. Pham, and J. Raynal, Z. Phys. A - Hadrons and Nuclei {\bf 343}, 179 (1992).
\bibitem{6} D. L. Pham and R. de Swiniarski, Nuovo Cimento A {\bf 107}, 1405 (1994).
\bibitem{7} J. J. Kelly, Phys. Rev. C {\bf71}, 064610 (2005).
\bibitem{8} S. Shim, M. W. Kim, B. C. Clark, and L. Kurth Kerr, Phys. Rev. C {\bf 59}, 317 (1999).
\bibitem{9} S. Shim, Shin-Ho Ryu and Min-Soo Kim, J. Korean. Phys. Soc. {\bf 51}, 271 (2007); S. Shim, Shin-Ho Ryu and Min-Soo Kim, J. Korean. Phys. Soc. {\bf 53}, 1146 (2008).
\bibitem{10} S. Shim and M. W. Kim, Int. Jou. of Mod. Phys. E {\bf 21}, 1250098 (2012).
\bibitem{11} J. Raynal, {\it Computing as a Language of Physics}, ICTP International Seminar Course, 281(IAEA, Italy, 1972); J. Raynal, {\it Notes on ECIS94}, Note CEA-N-2772, 1994.
\bibitem{12} C. J. Horowitz and B. D. Serot, Nucl. Phys. A {\bf 368}, 503 (1981).
\bibitem{13} R. J.  Furnstahl, C. E.  Price, and G.  E. Walker, Phys.  Rev. C  {\bf 36}, 2590 (1987).
\bibitem{14} L. Ray and G. W. Hoffmann, Phys. Rev. C {\bf 31}, 538 (1986).
\bibitem{15} S. Shim and M. W. Kim, J. Korean. Phys. Soc. {\bf 64}, 483 (2014).
\bibitem{16} G. S. Blanpied, B. G. Ritchie, M. L. Barlett, R. W. Fergerson, G. W. Hoffmann, J. A. McGill, B. H. Wildenthal, Phys. Rev. C {\bf 38}, 2180 (1988).
\bibitem{17} G. S. Blanpied, G. A. Balchin, G. E. Langston, B. G. Ritchie, M. L. Barlett, G. W. Hoffmann, J. A. McGill, M. A. Franey, M. Gazzaly, B. H. Wildenthal, Phys. Rev. C {\bf 30}, 1233 (1984).
\bibitem{18} G. S. Blanpied, B. G. Ritchie, M. L. Barlett, R. W. Fergerson, G. W. Hoffmann, J. A. McGill, B. H. Wildenthal, Phys. Rev. C {\bf 37}, 1987 (1988).
\bibitem{19} R. de Swiniarski, A. D. Bacher, F. G. Resmini, G. R. Plattner, D. L. Hendrie and J. Raynal, Phys. Rev. Lett. {\bf 28}, 1139 (1972).
\end{references}
\end{document}